\documentclass[12pt]{iopart}
\usepackage{iopams}
\usepackage{graphicx}
\usepackage{cite}
\expandafter\let\csname equation*\endcsname\relax
\expandafter\let\csname endequation*\endcsname\relax
\usepackage{amsmath}
\usepackage[colorlinks,
    linkcolor=blue,
    anchorcolor=blue,
    citecolor=blue
]{hyperref}

\begin{document}
	
\title{Characteristics of the edge temperature ring oscillation during stationary improved confinement mode in EAST}
\renewcommand{\thefootnote}{\fnsymbol{footnote}}
\author{A. D. Liu$^1$, X. L. Zou$^{2}$, X.M. Zhong$^{1\ast}$, Y.T. Song$^{3\ast}$, M.K. Han$^{4}$, Y.M. Duan$^{3}$, H.Q. Liu$^{3}$, T.B. Wang$^{6}$, E.Z. Li$^{3}$, L. Zhang$^{3}$, X. Feng$^{5}$, G. Zhuang$^{1}$ and EAST I-mode working group$^{\dagger}$}


\address{$^1$ School of Nuclear Science and Technology, University of Science and Technology of China, Anhui Hefei 230026, China}
\address{$^2$ CEA, IRFM, F-13108 St Paul Lez Durance, France}
\address{$^3$ Institute of Plasma Physics, Chinese Academy of Sciences, Anhui Hefei 230021, China}
\address{$^4$ Southwestern Institute of Physics, Chengdu 610041, China}
\address{$^5$ Advanced Energy Research Center, Shenzhen University, Shenzhen 518060, China}
\address{$^6$ ITER organization, 13115 Saint Paul Lez Durance, France}
\address{$\ast$ Authors to whom any correspondence should be addressed.}
\address{$\dagger$  See the appendix.}
\ead{zhongxm@mail.ustc.edu.cn, songyt@ipp.ac.cn}

\date{\today}

\begin{abstract}
I-mode is a natural ELMy-free regime with H-mode like improved energy confinement and L-mode like particle confinement, making it an attractive scenario for future tokamak based fusion reactors. A kind of low frequency oscillation was widely found and appeared to be unique in I-mode, with the frequency between stationary zonal flow and geodesic-acoustic mode (GAM) zonal flow. In EAST, $90$ percent I-mode shots have such mode, called edge temperature ring oscillation (ETRO). The mode probably plays an important role during I-mode development and sustainment, while investigations are needed to clarify the differences between ETRO and the similar mode named as low frequency edge oscillation (LFEO) in AUG and C-Mod, especially whether it is still GAM. In the paper, the ETRO characteristics in EAST were investigated in detail and most do not agree with GAM, including that 1) during L-I transition with edge $T_e$ and $T_i$ both increasing, ETRO has a smaller frequency than GAM; 2) ETRO has distinct harmonics in various diagnostics; 3) The magnetic component of ETRO is dominated by $m=1$ structure; 4) ETRO is accompanied by turbulence transition between electron-scale and ion-scale; 5) As I-mode approaching to H-mode, ETRO frequency would decrease rapidly with $T_e$ increasing. These features imply that ETRO is probably caused by the stationary zonal flow with finite frequency. Moreover, other damping mechanisms need to be involved besides collision in the I-mode edge region. It was found that modest fueling could decrease the ETRO intensity with the I-mode confinement sustaining, suggesting that supersonic molecular beam injection (SMBI) could be used as an effective tool to control ETRO.
\end{abstract}



\maketitle

\section{Introduction}
Although high confinement mode (H-mode) \cite{Wagner1984PRL} has been considered as the baseline operation scenario for the International Thermonuclear Experimental Reactor (ITER) \cite{rebut1995}, the heat load caused by large edge localized modes (ELMs) due to relaxation of edge pressure and current is still one of the most crucial issues in fusion research \cite{Loarte2003JNM}. An alternative improved confinement regime (I-mode), featuring high energy confinement comparable to H-mode and moderate particle confinement comparable to L-mode \cite{Whyte2010NF}, may be a possible solution and has been widely investigated on various divertor Tokamaks \cite{Hubbard2016NF,Ryter2017NF,Happel2019NME,Feng2019NF,Liang2023NF}. Generally, I-mode is usually obtained under the unfavorable configuration, i.e. the  $B$$\times$$\nabla$$B$ ion drift direction pointing away from the X-point. In C-Mod and AUG, I-mode is always accompanied by the weakly coherent mode (WCM) and the geodesic-acoustic mode (GAM), moreover, GAM may play an important role during L-I-H transitions through nonlinear flow-turbulence couplings \cite{Cziegler2017PRL}. While in EAST, a low-frequency coherent mode, which was identified as a radially localized edge temperature ring oscillation with azimuthally symmetric structure in the pedestal region, named as ETRO, was reported during stationary I-mode \cite{Feng2019NF,Liu_2020}. The frequency of ETRO is between the limit cycle oscillating (LCO) and GAM in EAST, implying that ETRO is probably neither the stationary zonal flows triggering marginal L-mode to H-mode transition \cite{XuPRL2011}, nor the ordinary GAM zonal flows \cite{Conway2021NF}. Recently, a similar mode called low frequency edge oscillation (LFEO) were also been reported in C-Mode and AUG \cite{Mccarthy2022PHD,Bielajew2022POP}, however, LFEO could be found in only $40\%-60\%$ I-mode shots and some features are different with ETRO. While LFEO is regarded as a GAM \cite{Mccarthy2022PHD}. Considering that more than $90\%$ I-mode shots in EAST are accompanied by ETRO and it probably played important role for sustaining the stationary I-mode in EAST, the characteristics of ETRO, especially the differences with GAM were systematically investigated in the paper.

The remainder of this paper is organized as follows. The typical stationary I-mode shots and how ETRO emerged are presented in section II. In section III, structures of ETRO are displayed, including the symmetrical temperature components and asymmetric magnetic components. The turbulence transition accompanied by ETRO and the linear simulation results are shown in section IV. The statistics of ETRO frequency is given in section V, as well as the ETRO evolution during I-mode to H-mode transition. The effect of SMBI on ETRO, edge plasma flow and turbulences is displayed in section VI. Finally, a summary on the ETRO charateristics and the comparison with GAM is made in Section VII.

\section{Stationary I-mode in EAST}

\begin{figure}[htbp]
	\centering
	\includegraphics[width=0.8\textwidth,clip]{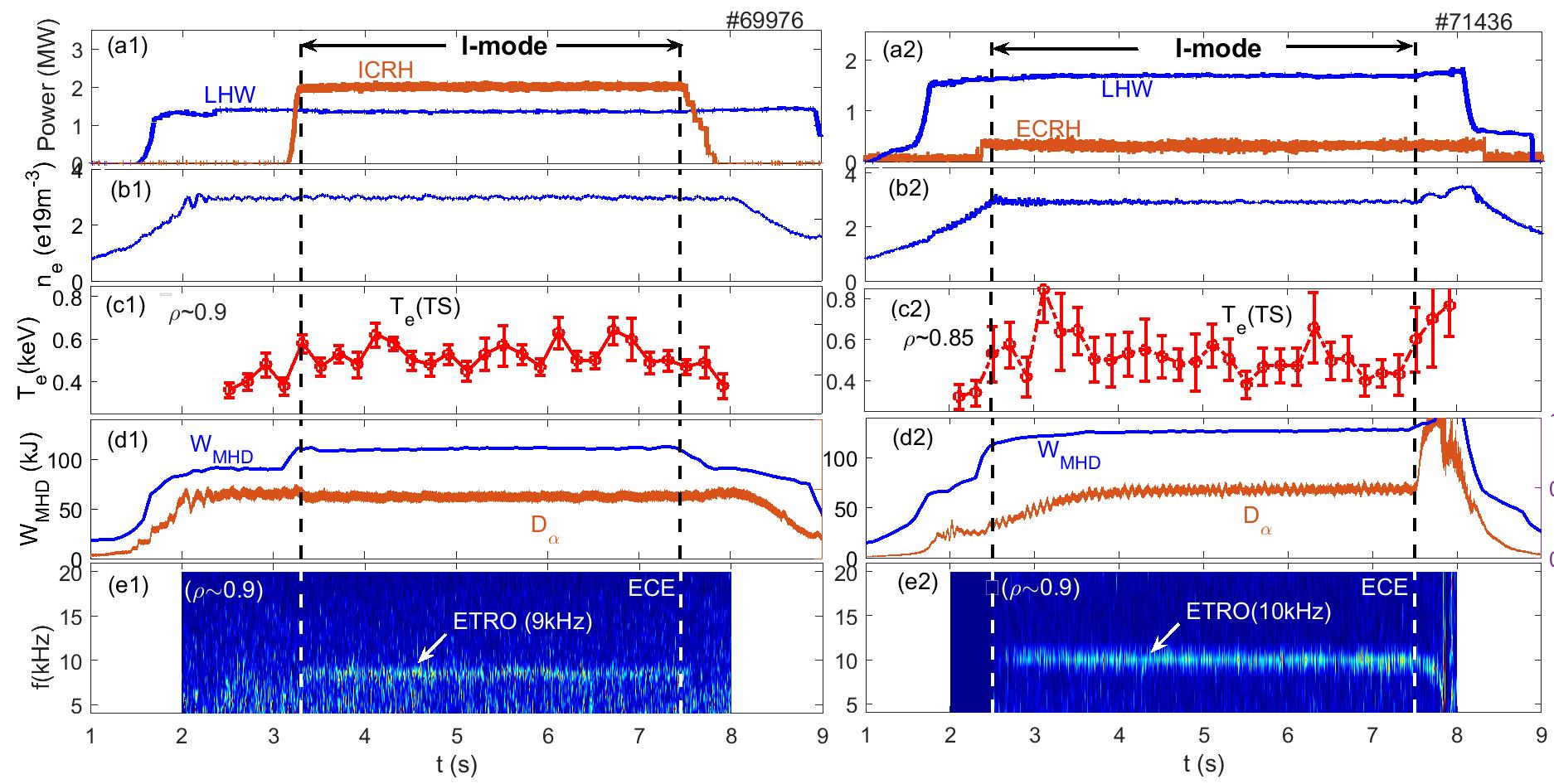}
	\caption{\label{f1} From top to bottom, temporal evolutions of (a) auxiliary heating power, (b) chord averaged density, (c) edge $T_e$, (d) plasma stored energy $W_{MHD}$ and $D_{\alpha}$ signal, and (e) time-frequency spectrum of the edge ECE signal in typical stationary I-mode shot $69976$ and $71436$.}
	
\end{figure}

The most noteworthy feature of I-mode in EAST is that it could spontaneously last for several seconds. Figure \ref{f1} showed two typical stationary I-mode discharges $69976$ and $71436$. From the top to bottom in fig.\ref{f1}, the curves gave the auxiliary heating power, the chord-averaged density, the edge $T_e$ evolution from the Thomson scattering system, the plasma stored energy $W_{MHD}$, the $D_{\alpha}$ signal, and the time-frequency spectrum from edge ECE channel. The auxiliary heating in $69976$ included $1.4MW$ low hybrid wave (LHW) and $2MW$ ion cyclotron resonance heating (ICRH), and $W_{MHD}$ was increased to a new plateau with no distinct change in $D_{\alpha}$ signal during the transition. The central frequency of ETRO are $8-9kHz$, which appeared during I-mode confinement. The auxiliary heating in $71436$ included $1.5MW$ low hybrid wave (LHW) and $0.5MW$ electron cyclotron resonance heating (ECRH), and as $W_{MHD}$ increasing to a new plateau $D_{\alpha}$ signal was continuously increased during the transition. The central frequency of ETRO are $10kHz$. The plasma entered into H-mode at about $7.8s$ in $71436$.


\begin{figure}[htbp]
	\centering
	\includegraphics[width=0.5\textwidth]{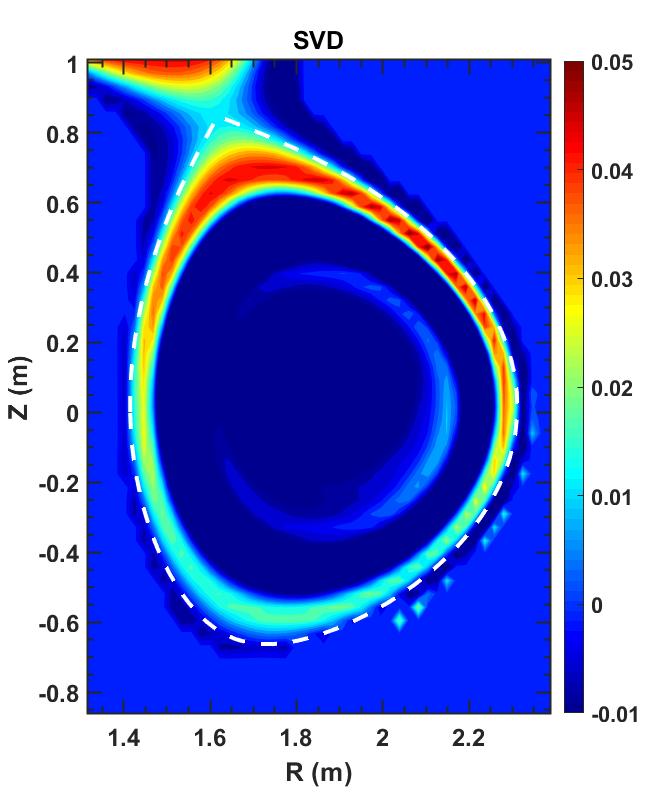}
	\caption{\label{ETROSVD} the second singular value decomposition (SVD) result of tomography reconstruction through the 64-channel bolometer arrays.}
\end{figure}

The origin of ETRO definition has been presented in our previous paper in detail \cite{Liu_2020}, which is an azimuthally symmetry and radially localized structure dominated by electron temperature fluctuions.
Fig. \ref{ETROSVD} shows the typical poloidal structure estimated through tomography reconstruction based on the 64-channel bolometer signals. It is clearly observed that ETRO is an in-phase ring-like structure localized around $\rho = 0.9$ with a small up-down amplitude asymmetry.

\begin{figure}[htbp]
\centering
	\includegraphics[width=0.8\textwidth,clip]{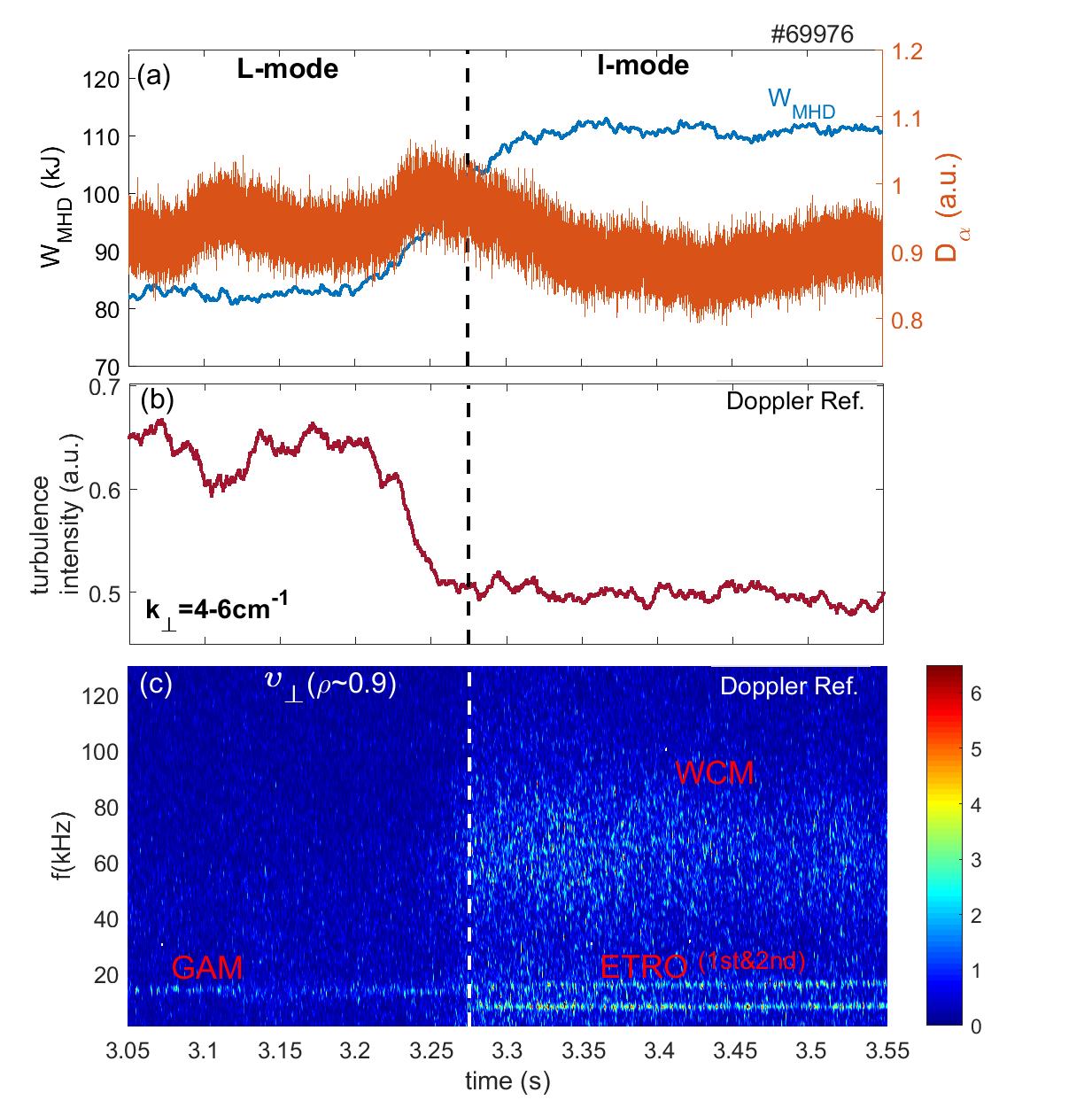}
	\caption{\label{f2} (a) Plasma stored energy $W_{MHD}$ and $D_{\alpha}$, (b) turbulence intensity, and (c) time-frequency spectrum of turbulence rotation velocity $\upsilon_{\perp}$ at $\rho\sim0.9$ from Doppler reflectometry during L-mode to I-mode transition.}
\end{figure}

Evolutions of the edge turbulence intensity, as well as the rotation velocity ($\upsilon_{\perp}$) measured through the multi-channel Doppler reflectometry (DR) \cite{Zhou2013RSI,Hu2017RSI} during L-mode to I-mode transition were shown in fig.\ref{f2}. The measured perpendicular wavenumber $k_\perp$ at the cutoff layer is about $4 - 6 cm^{-1}$. In shot $69976$, it could be seen that distinct WCM and ETRO appeared simultaneously after $t=3.27s$, before that time GAM could also be found, with a smaller intensity compared to ETRO. Meanwhile the turbulence intensity would be significantly decreased. It could be noted that second harmonics of ETRO could also be seen directly.

\begin{figure}[htbp]
\centering
	\includegraphics[width=0.8\textwidth,clip]{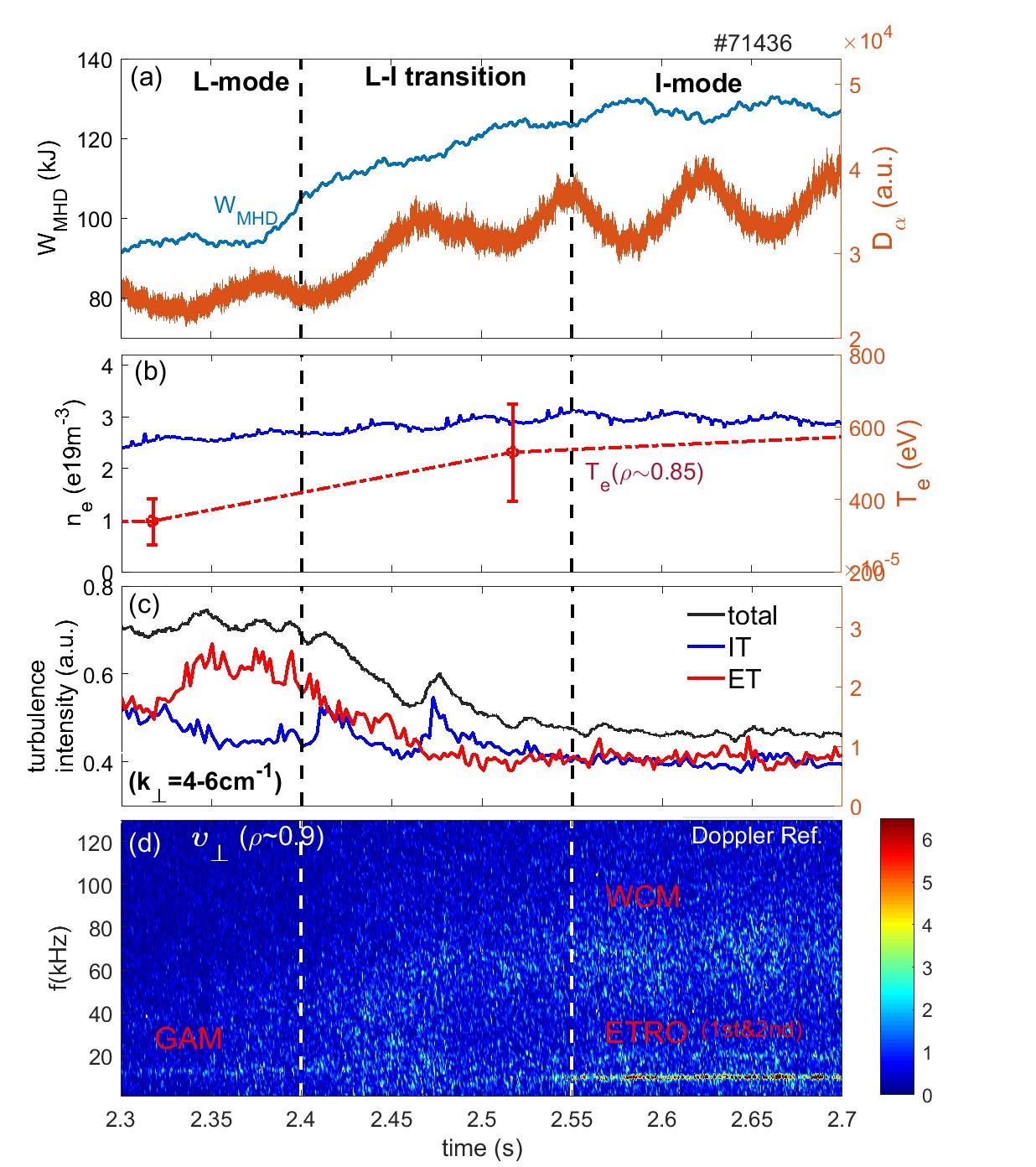}
	\caption{\label{f3} (a) Plasma stored energy $W_{MHD}$ and $D_{\alpha}$ (b) chord averaged density and edge $T_e$, (c) turbulence intensity, and (d) time-frequency spectrum of turbulence rotation velocity $\upsilon_{\perp}$ at $\rho\sim0.9$ during L-mode to I-mode transition in shot $71436$.}
\end{figure}

Another typical type of L-I transition was shown in fig.\ref{f3}. In shot $71436$, the L-I transition seemed much slower. From $t=2.4s$ to $t=2.55s$, while the slow transition is consistent with the slow decay of the total turbulence intensity shown in fig.\ref{f3}(c). The rebound at about $t=2.47s$ further delayed the descent time, and the rebound is mainly contributed from the turbulence at ion diamagnetic drift direction (IT). Here ET means turbulence at electron diamagnetic drift direction and the calculation of IT and ET will be further described in section $4$. This panel implied that suppression IT seems to be the primary prerequisite for entering I-mode. From fig.\ref{f3}(d) it could be seen that the low frequency components of $\upsilon_{\perp}$ spectrum seemed chaotic during the slow transition stage, while GAM, ETRO and WCM may all exist. These phenomena are probably due to the marginal injection heating power in this shot, which has only $1.5MW$ LHW and $0.5MW$ ECRH. 

\section{features of ETRO structure}

\begin{figure}[htbp]
\centering
	\includegraphics[width=0.8\textwidth,clip]{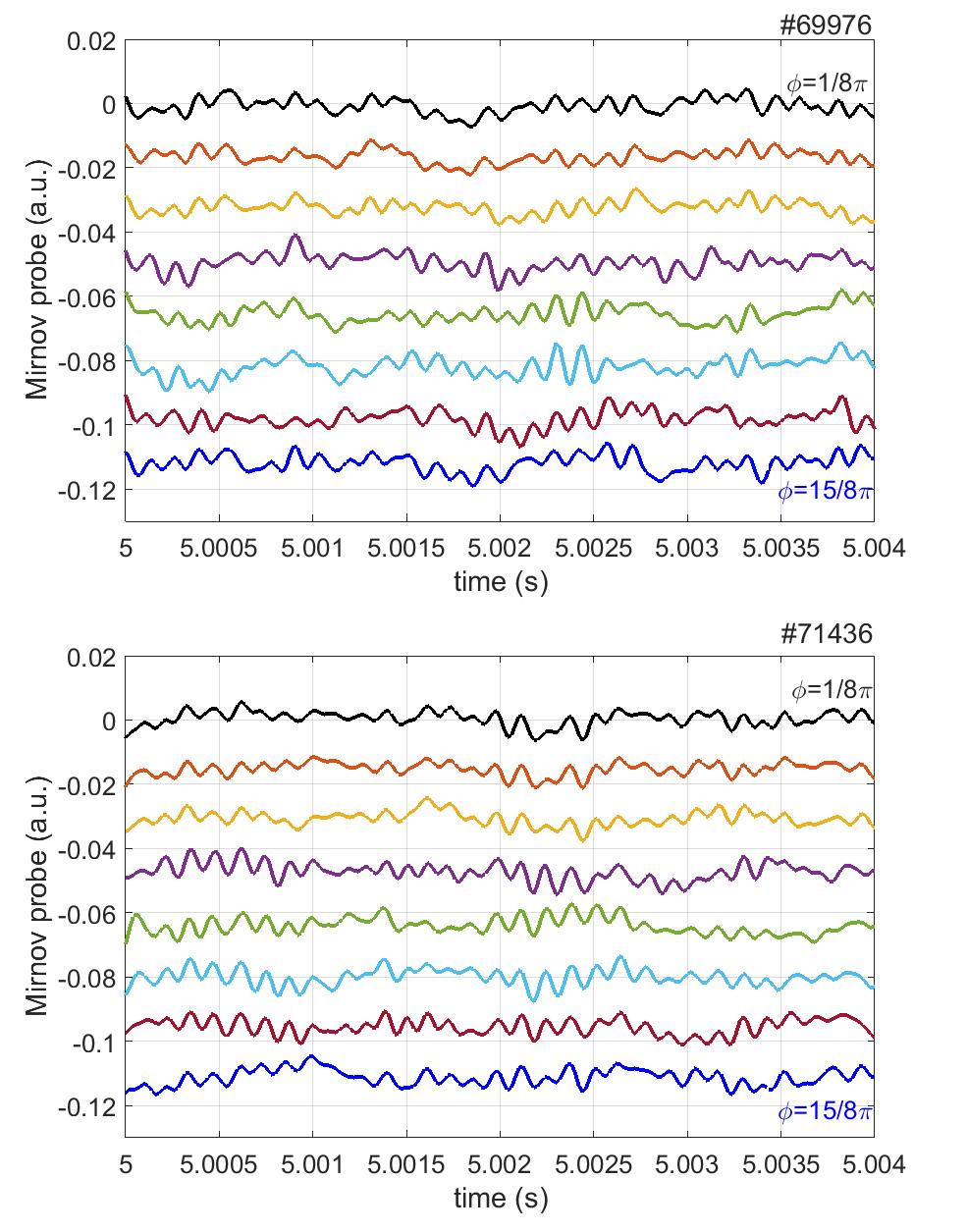}
	\caption{\label{f4} The original signals from $8$ toroidal magnetic probes in shot $69976$ and $71436$ respectively. (The averaged values were vertically shifted for clarify)}
\end{figure}

ETRO could be found in most edge plasma diagnostics, such as magnetic coils, divertor probes, $D_{\alpha}$ filterscopes, electron cyclotron emissions (ECE), bolometers, soft-X ray, Doppler reflectometry, and sometimes the polarimeter-interferometer (POINT) \cite{Liu2016RSI}. The toroidal symmetry could be directly identified from $16$ groups of toroidal magnetic probes. In fig.\ref{f4} the original $4$ millisecond signals from $8$ magnetic probes during I-mode are shown. 

\begin{figure}[htbp]
\centering
	\includegraphics[width=0.8\textwidth,clip]{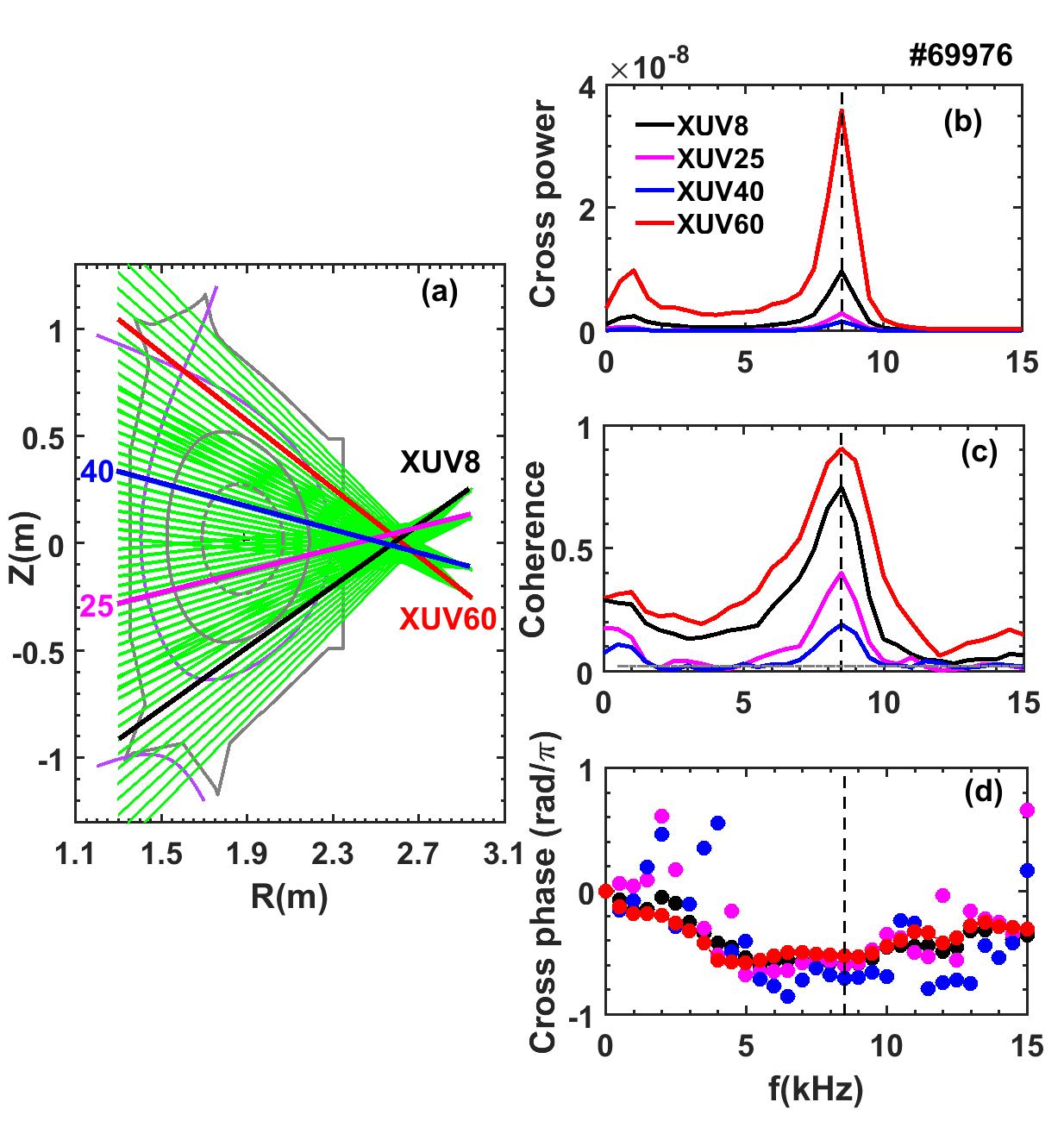}
	\caption{\label{f5} (a) Arrangement of $64$-channel XUV array, the frequency spectra of (b) cross-power, (c) coefficient and (d) cross-phase between $4$ XUV signals and the DR signal during I-mode.}
\end{figure}

The poloidal symmetry of ETRO in fig.\ref{ETROSVD} is estimated through radiation reconstruction delivered by Gaussian process tomography algorithm \cite{Wang2018RSI}. Actually, the poloidal symmetry could also be estimated from the coherence analyses between DR signal and the bolometers, as shown in fig.\ref{f5}. Distribution of the $64$ XUV array in the cross section is shown in fig.\ref{f5}(a), and there are $4$ groups with $16$ channels for each group, the detectors are all located around the middle plane \cite{Duan2011PST}. It could be seen that channel $XUV8$ and channel $XUV60$ would pass through the I-mode pedestal region from top and bottom respectively. Figure \ref{f5}(b-d) showed the spectra of cross-power, coefficient, and cross-phase between $4$ bolometer signals ($XUV8,25,40,60$) and one reference DR signal. It could be found that all spectra have distinct peaks at the ETRO frequency and the coefficients of channel $XUV8$ and $XUV60$ are the two maximum, while $XUV60$ has the largest intensity and coherent coefficient because it is close to the X-point. Moreover, the cross-phases at the ETRO frequency are similar for all channels, suggesting that ETRO fluctuation has symmetry phases along the flux surface while the fluctuation amplitude reached maximums around the top and bottom, consistent with the tomography reconstruction results \cite{Liu_2020}.

\begin{figure}[htbp]
	\centering
	\includegraphics[width=0.6\textwidth]{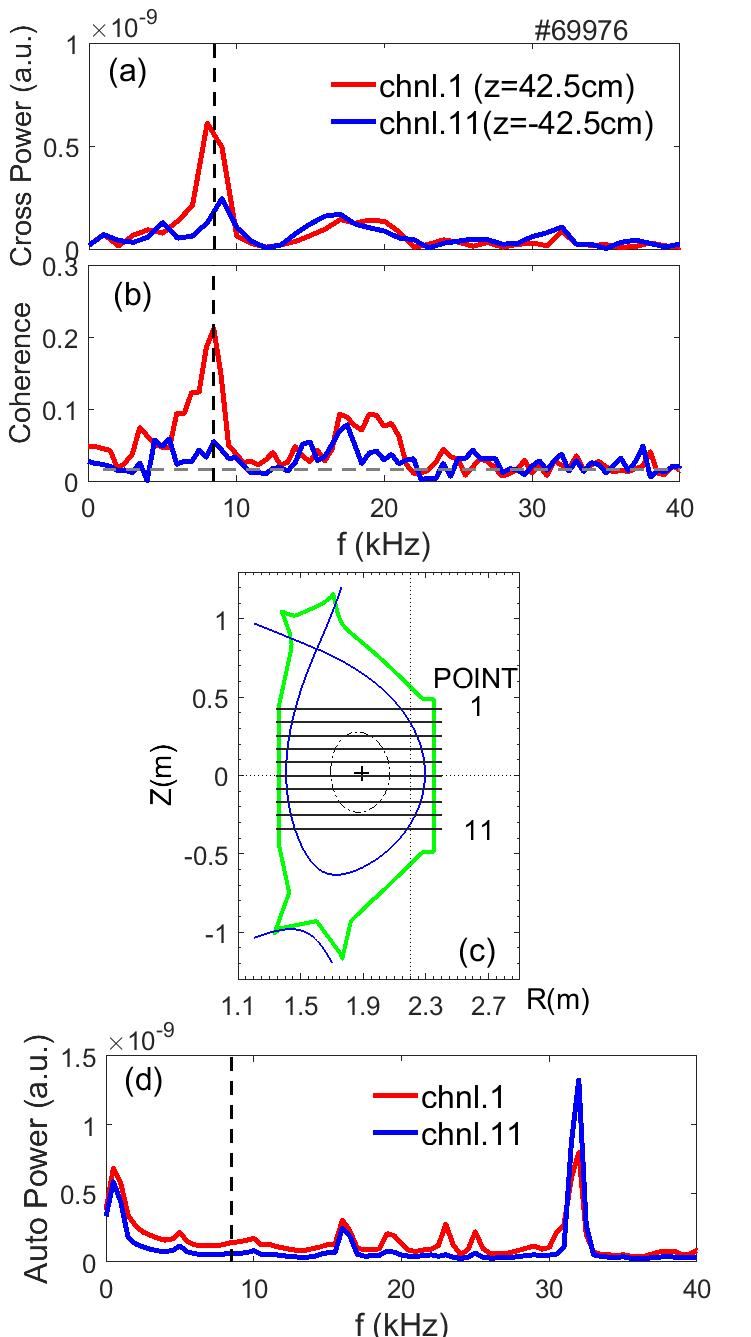}\\
	\caption{\label{f6} (a) Cross power and (b) coefficient spectra between POINT channel $1$ and $11$ and DR signal during I-mode. (c) Arrangement of POINT diagnostic. (d) ETRO didn't appear in the auto power spectra of POINT signals.}
\end{figure}

A question is whether ETRO has density fluctuations. Figure \ref{f6} showed the coefficient spectra between two channels from POINT and the reference DR signal used in fig.\ref{f5}. POINT in EAST has $11$ chord-averaged horizontal channel \cite{Liu2016RSI}, while channel $1$ is the top chord at $Z=42.5cm$ and channel $11$ is the bottom chord at $Z=-42.5cm$, as shown in the bottom panel in fig.\ref{f6}. It is not surprising that distinct coherence could also be found at the ETRO frequency for the two channels, suggesting that ETRO still has distinguishable density components. Moreover, similar to the poloidal distribution of temperature perturbations, the density perturbations caused by ETRO also reached maximum around the up and down X-point. However, the coherent coefficient of density fluctuation is much smaller than that of temperature fluctuation shown in fig.\ref{f5}. Considering that the ETRO peak could not be estimated directly from the self power spectrum of POINT signal, the relative density perturbation amplitude of ETRO should be below $2\%$ according to the phase noise level of POINT system \cite{Liu2016RSI}. 

\begin{figure}[htbp]
\centering
	\includegraphics[width=0.8\textwidth]{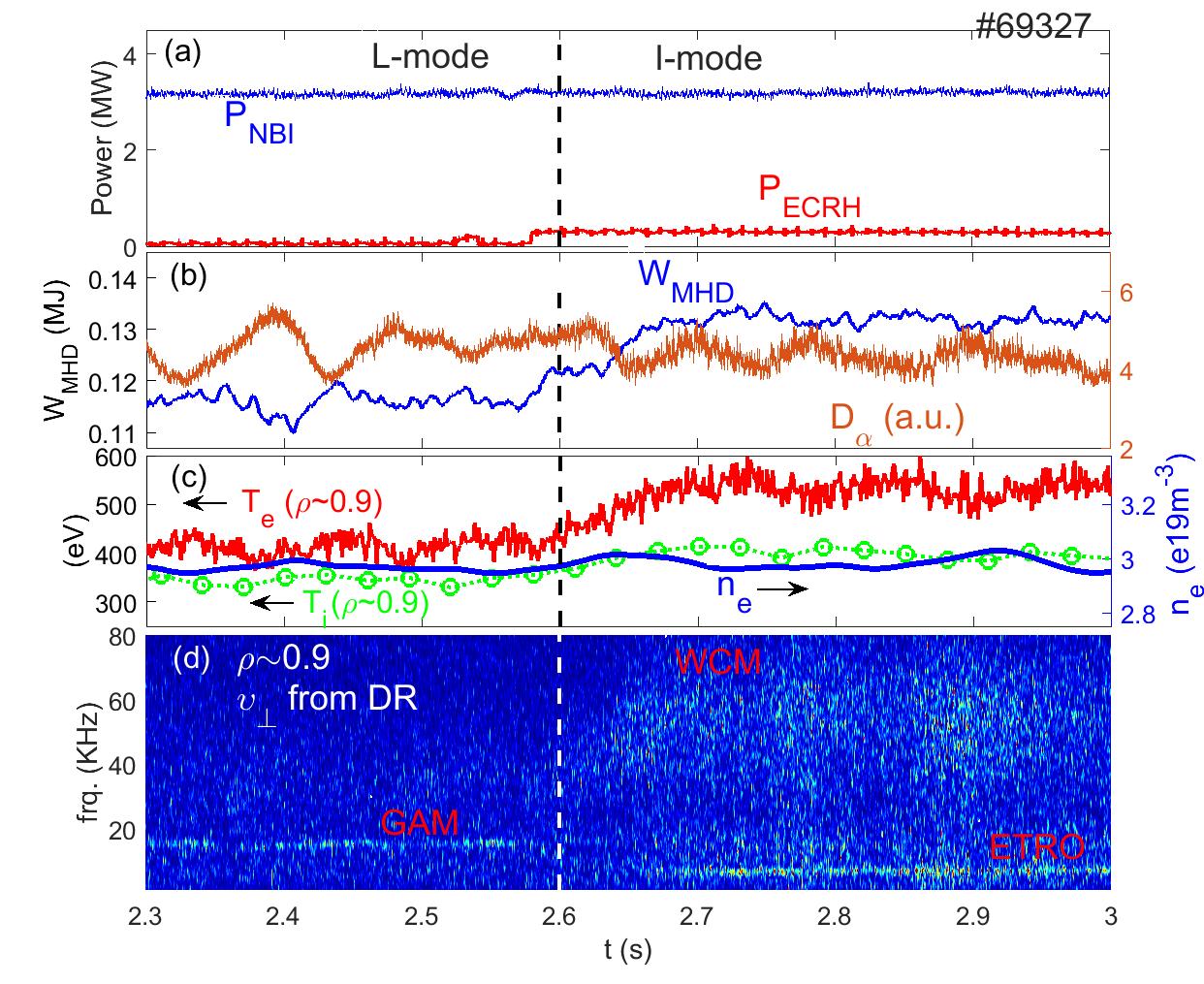}\\
	\caption{\label{f7} From top to bottom, temporal evolutions of auxiliary heating power, plasma stored energy $W_{MHD}$ and $D_{\alpha}$ signal, chord averaged density, edge $T_e$ and $T_i$, and time-frequency spectrum of turbulence rotation velocity $\upsilon_{\perp}$ during L-I transition in shot $69327$.}
\end{figure}

The azimuthally symmetric structure of ETRO strongly implied that it is probably a kind of zonal field or zonal flow. In toroidal plasmas, two kinds of zonal flows exist \cite{Diamond2005PPCF}. One is low frequency (or stationary) zonal flow \cite{Fujisawa2004PRL}, and the other is geodesic acoustic mode (GAM) \cite{Itoh2005PPCF,Conway2021NF}. Considering that the ETRO frequency is about $6-12kHz$, it seems that ETRO may belong to the GAM type. However, much experimental evidence indicated that ETRO is not the typical GAM. Figure \ref{f7} showed an L-I transition without low hybrid wave heating. In this shot, the electron temperature from ECE and the ion temperature from CXRS \cite{Li2014RSI} in the edge region are both measured. It could be found that during L-I transition both $T_e$ and $T_i$ increased while chord-averaged $n_e$ kept nearly unchanged. The measurement location of DR edge channel \cite{Hu2017RSI} only depended on the density profile and the magnetic field, which is also unchanged during L-I transition. In L-mode, the frequency of GAM is about $18kHz$, according to the dispersion relationship:  $\omega_{GAM}\simeq\sqrt{2}C_s/R$, where $R$ and $C_s$ are the major radius and the ion sound velocity, respectively, then GAM frequency will definitely increase in I-mode. However, ETRO frequency is just $7-8kHz$ in this shot, suggesting that ETRO would not be GAM. 

\begin{figure}[htbp]
\centering
	\includegraphics[width=0.8\textwidth]{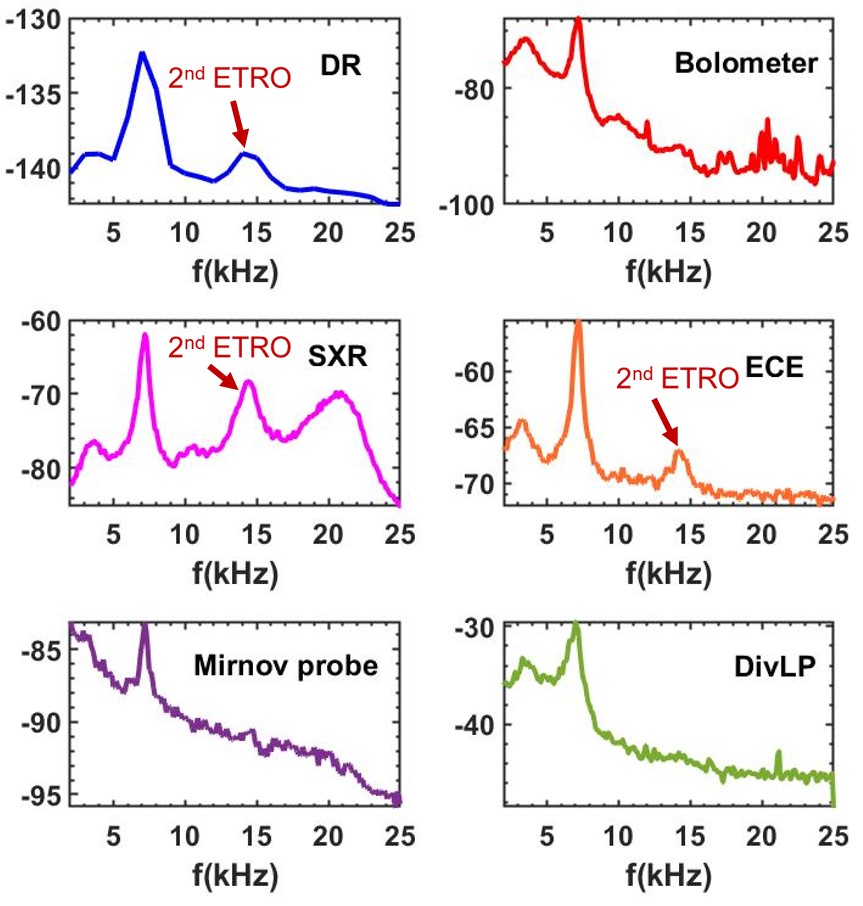}\\
	\caption{\label{f8} The power spectra of signals from DR, Bolometer, Soft X-ray, ECE, Mirnov probe and Divertor Langmuir probe in I-mode shot $69327$.}
\end{figure}

Another evidence is the second harmonics of ETRO, as shown in fig.\ref{f8}, where the peaks at $14-15kHz$ could be found from the spectra of DR phase, Soft X-ray, and ECE spectra. As mentioned above, just based on the central frequency values, it could be concluded that these peaks should be the second harmonics of ETRO. In our previous paper \cite{Liu_2020}, the ETRO harmonics in the DR phase signal could be well explained as the amplitude asymmetry of turbulence rotation velocity $\upsilon_{\perp}$ resulted from the alternating turbulence transitions:  $\vec{\upsilon}_{\perp}=\vec{\upsilon}_{E\times B}+\vec{\upsilon}_{pha}$, where $\vec{\upsilon}_{E\times B}$ is the plasma $E\times B$ rotation and $\vec{\upsilon}_{pha}$ is the turbulence phase velocity. While the second harmonics of GAM were observed only in the bicoherence analyses \cite{Nagashima2007PPCF}, or for the GAM driven by the energetic particle \cite{Horvath2016NF,Qiu2017POP}, further suggesting that ETRO would not be GAM. 

\begin{figure}[htbp]
\centering
	\includegraphics[width=0.8\textwidth,clip]{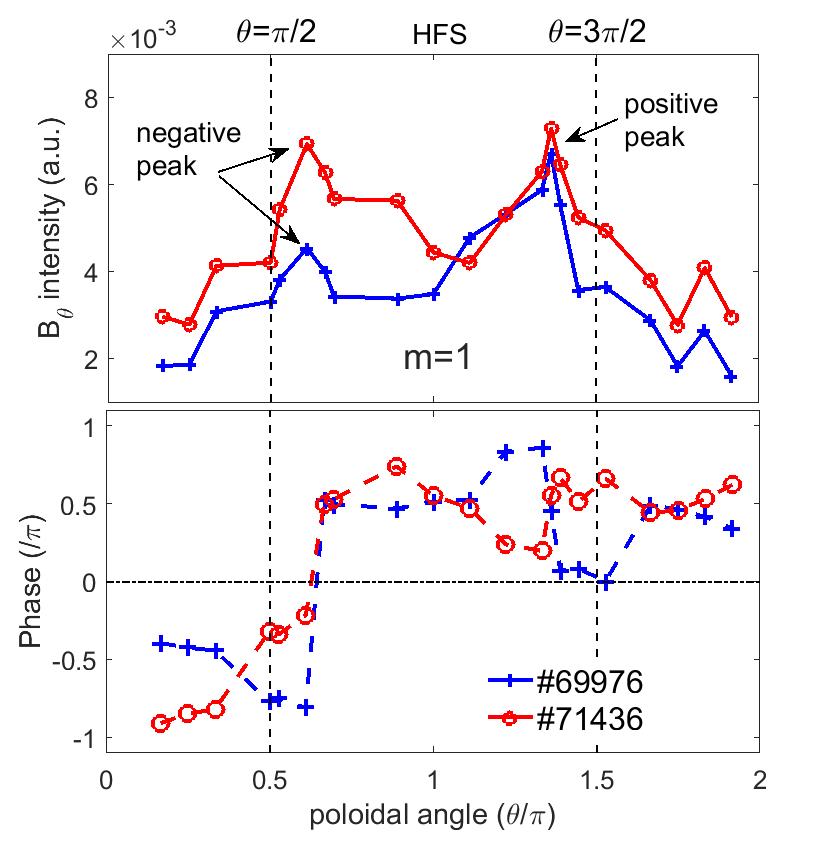}\\
	\caption{\label{f9} The poloidal distributions of (a) intensities and (b) phases of magnetic fluctuation at the ETRO frequency in shot $69976$ and $71436$.}
\end{figure}

The poloidal distribution of magnetic components of ETRO is shown in fig.\ref{f9} through coherence analyses between poloidal magnetic probes and one reference DR signal. The intensities and phases along poloidal direction both indicated that the poloidal magnetic components of ETRO are dominated by a twisted $m=1$ structure, and the intensity at high field side (HFS) is much lower than that at low field side (LFS). Moreover, the phase folding taking place around $\theta\sim1.3\pi$ further suggested that other harmonic components are present as well \cite{Kim2001PPCF}. It should be noted that both theories and experiments point out that the magnetic component of GAM is dominated by $m=2$ structure \cite{Conway2021NF}. 

\begin{figure}[htbp]
\centering
	\includegraphics[width=0.8\textwidth,clip]{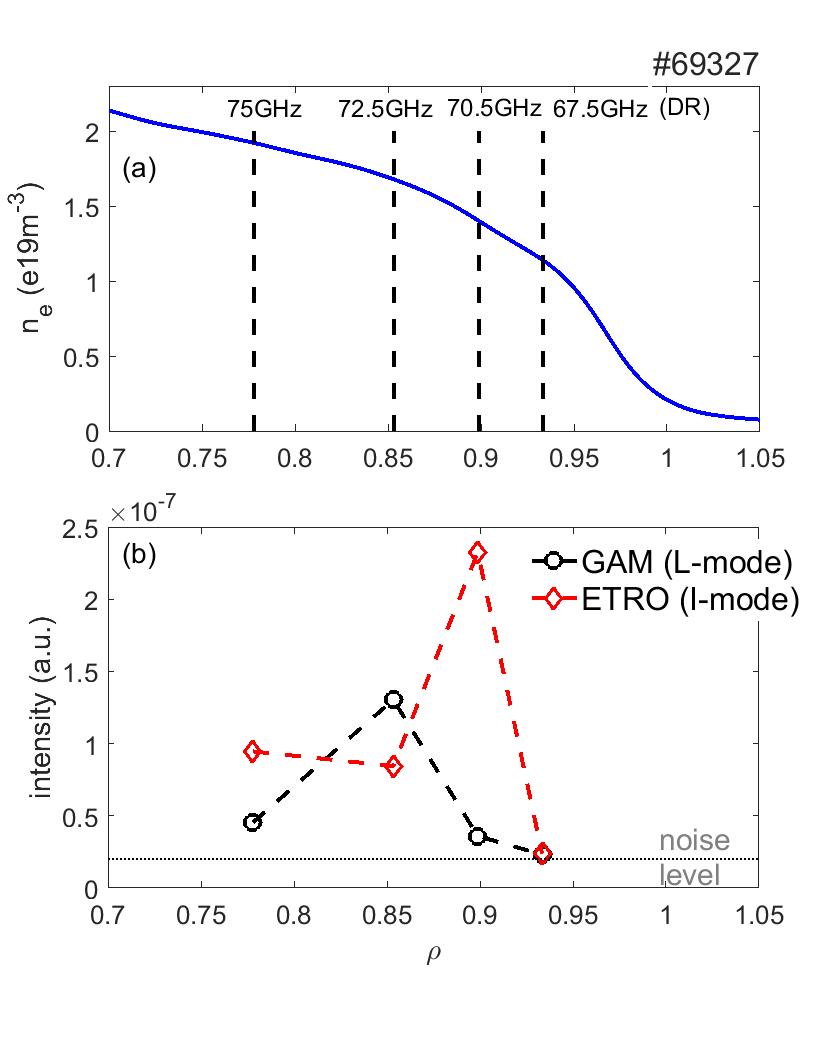}\\
	\caption{\label{f10} (a) The density profile and the measurement positions of multi-channel DR, with (b) the radial distributions of GAM intensity during L-mode and ETRO intensity during I-mode. }
\end{figure}

In shot $69327$, there are four DR channels located inside the LCFS, as indicated by the vertical dash lines in fig.\ref{f10}(a), then the radial distributions of GAM and ETRO intensity could be estimated from the DR phase signals, as shown in fig.\ref{f10}(b). It could be found that ETRO intensity decayed radially outward much more quickly than inward, which is the general feature of ETRO and should be closely related to the turbulent transitions at this location \cite{Liu_2020}.  The turbulence transition will be further described in the next section. On the other hand, the outward decay of GAM at the edge plasmas would be much gentler because that GAM propagating prefer outward than inward, which has been widely reported in both experiments and simulations \cite{Conway2021NF,Kong2013NF,Kong2017NF}.

\section{Turbulence transition accompanied by ETRO}
\begin{figure}[htbp]
\centering
	\includegraphics[width=0.8\textwidth,clip]{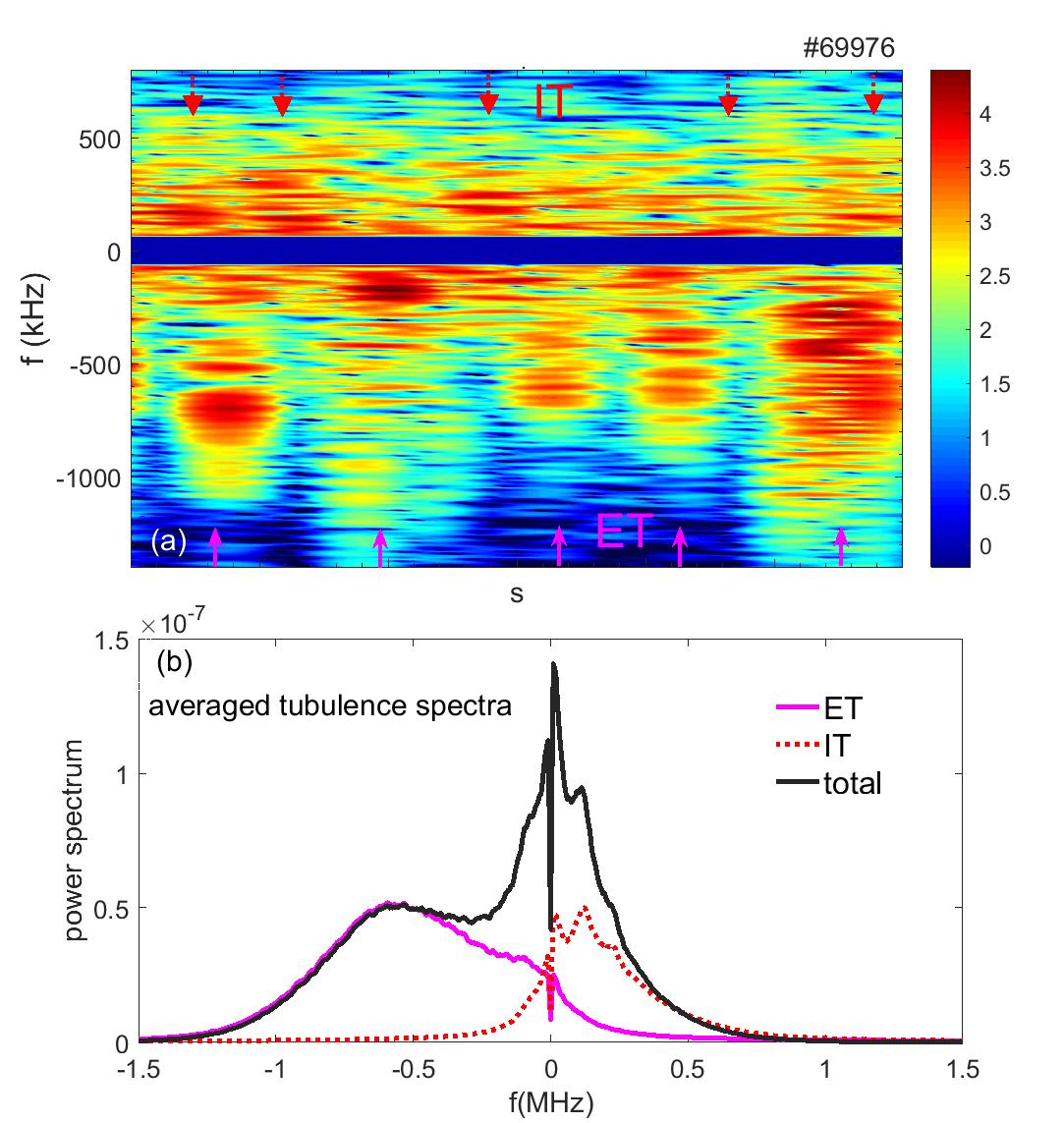}\\
	\caption{\label{f11} Temporal evolutions of turbulence rotation velocity $\upsilon_{\perp}$ (while curve in the middle) and turbulence spectra during $0.6ms$ I-mode in shot $69976$. ET means turbulence in electron diamagnetic drift direction while IT means turbulence in ion diamagnetic drift direction. }
\end{figure}

In the previous paper \cite{Liu_2020}, the alternating turbulence transitions accompanied by ETRO have been displayed. Here the turbulence and rotation velocity $\upsilon_{\perp}$ evolutions during $0.6$ms I-mode were shown in fig.\ref{f11}. In the figure, the while curves are $\upsilon_{\perp}$ fluctuations estimated from DR phase signals, while the contours are the time-frequency spectra of turbulence. It could be found that when turbulence in the frequency range of $[-800kHz\ -200kHz]$ (electron diamagnetic drift direction) appeared, $\upsilon_{\perp}$ became more negative, while when turbulence in the frequency range of $[200kHz\ 600kHz]$ (ion diamagnetic drift direction) appeared, $\upsilon_{\perp}$ became positive. In the following, the two kinds of turbulences are named as ET and IT respectively for simplicity, which were indicated by different arrows in fig. \ref{f10}. From the principle of DR measurement $\upsilon_{\perp}=\upsilon_{E\times B}+\upsilon_{pha}$, it could be concluded that assuming that $\upsilon_{E\times B}$ is constant, $\upsilon_{\perp}$ would have different fluctuation amplitude as ET and IT respectively dominated, well explaining why $\upsilon_{\perp}$ always has harmonics. While, experimentally the most common interaction between GAM and turbulence is the envelope modulation, which has been widely reported \cite{Conway2021NF} and is much different from the above phenomena.


\begin{figure}[htbp]
\centering
	\includegraphics[width=0.8\textwidth,clip]{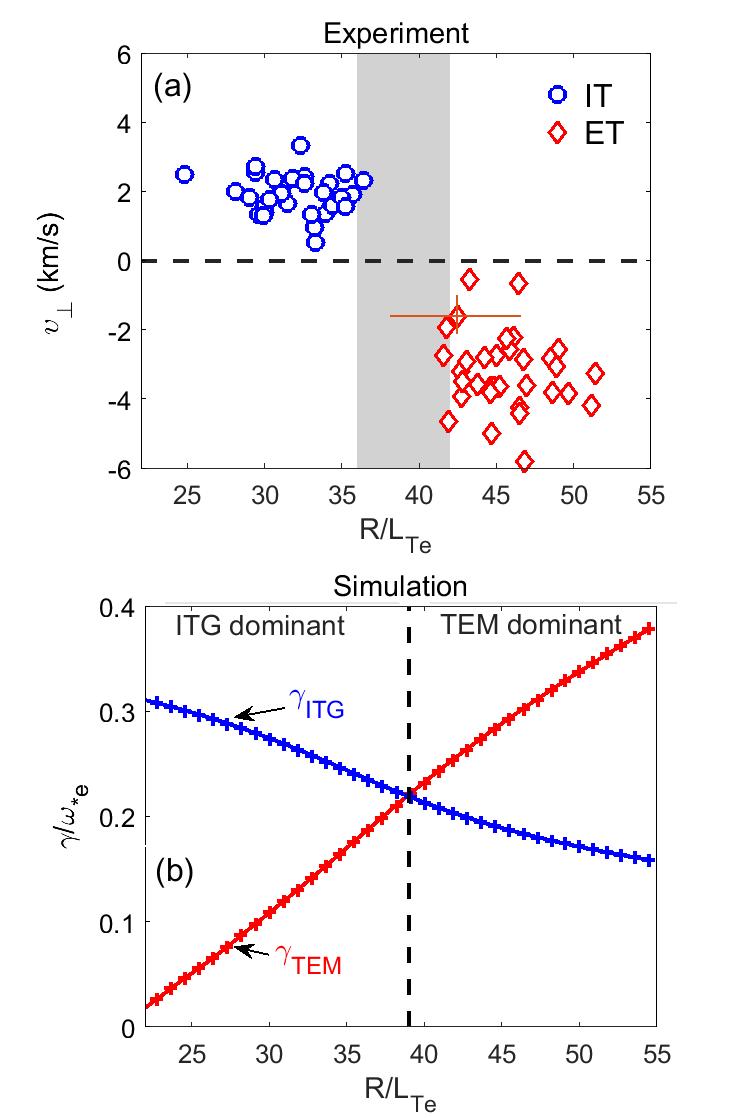}\\
	\caption{\label{f12} (a) $\upsilon_{\perp}$ versus $R/L_{T_e}$ from experimental data; (b) the normalized linear growth rate ($\gamma/\omega^{*}_e$) of TEM and ITG versus $R/L_{T_e}$ from HD7 gyrokinetic simulation.}
\end{figure}

Fig.\ref{f12}(a) plots the extreme values of $\upsilon_{\perp}$ as a function of $R/L_{T_e}$ during ETRO evolution. Considering that the detected turbulence wavenumber is $k_{\perp}=4-6cm^{-1}(k_{\perp}\rho_{s}=1.2-1.9)$ and $\eta_{e}=L_{n_e}/L_{T_e}$ increased much during L-I transition, TEM instabilities could be driven at the I-mode edge region by temperature gradient as $\eta_{e}>1$ \cite{Nilsson1995NF,Ryter2001PPCF}. On the other hand, considering that $\nabla T_{i}\simeq\nabla T_{e}$ is tenable \cite{Ryter2017NF}, ITG could also be stimulated when $\eta_{i}>1$. The toroidal gyrokinetic eigenvalue code HD7 \cite{Dong1995POP,Han2017NF} was used to calculate the linear growth rates of ITG and TEM with the following experimental parameters:  $R/L_{n_e}=R/L_{n_i}=9$, $R/L_{T_i}=20$. The results in Fig.\ref{f6}(b) showed that the linear growth rates of ITG and TEM have a crossover around $R/L_{T_e}=39$, implying that ITG-TEM transition may occur at this value, consistent with the experimental results.

Based on the simulation, ET is probably TEM and IT is probably ITG, and a simplified predator-prey model could be concluded to describe the ETRO cycle: as $\nabla T_e$ increased above the threshold, then dominant turbulence changed from ITG to TEM, and the outward particle flux suddenly increased and then decreased $T_e$ at the pedestal top, resulting in the dominant turbulence changing back to ITG. The cycle chain could keep $\nabla T_e$ oscillating around the marginal threshold and then maintain the I-mode confinement for a long time. It should be noted that the reduction of ITG accompanying with intermediate scale turbulence development was also reported in the gyro-fluid simulation for I-mode \cite{Manz2020NF}.

\section{center frequency of ETRO}

\begin{figure}[htbp]
\centering
	\includegraphics[width=0.8\textwidth,clip]{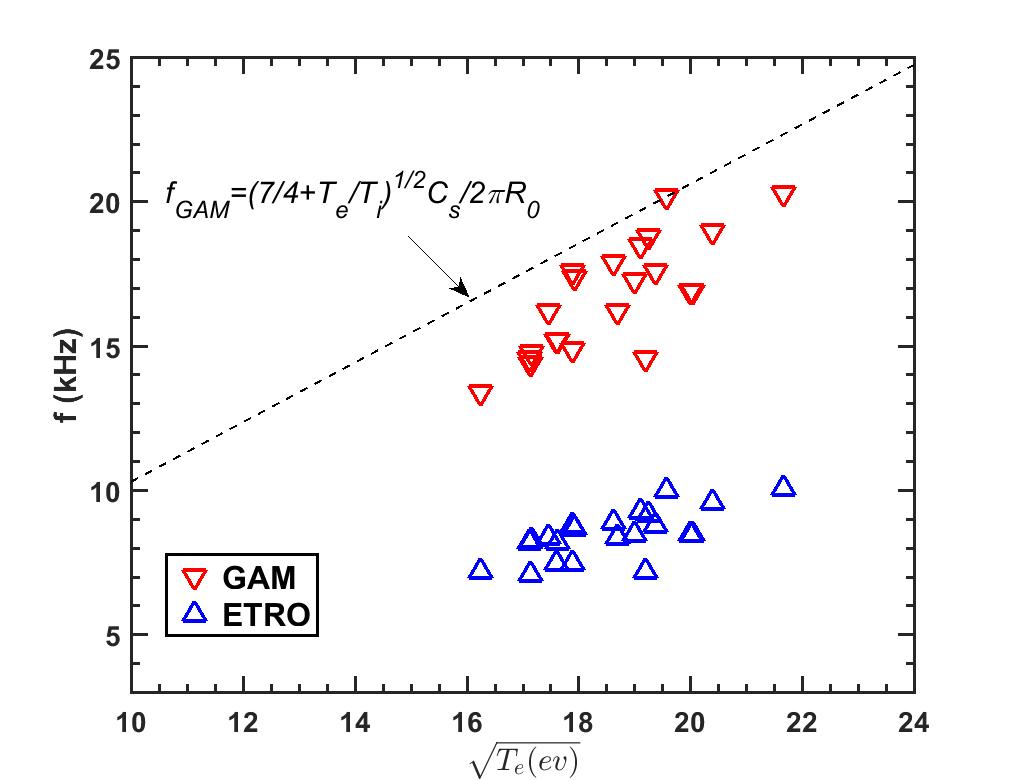}\\
	\caption{\label{f13} The central frequencies of GAM and ETRO versus squared electron temperature at the pedestal top in $20$ I-mode discharges with GAM and ETRO coexistence.}
\end{figure}

Considering that although GAM would always disappear at the location where ETRO is generated, it could still be detected at other radial positions. Based on the I-mode discharges with ETRO and GAM coexistence, the central frequencies of ETRO and GAM as a function of squared electron temperature at the pedestal top are displayed in fig.\ref{f13}, as well as the theoretical values of GAM frequency based on the formula: $f_{GAM}=\sqrt{7/4+T_e/T_i}Cs/(2\pi R_0)$ \cite{Conway2021NF}. Even though the electron temperature used here would be a little larger than that at the GAM generation location, the proportional relationship between GAM frequency and $C_s$ is still distinct. Moreover, it seems that ETRO frequency has a similar trend with GAM, suggesting that ETRO may still be the category of ion acoustic model. Considering that ETRO frequency is about $2-3$ times smaller than that of GAM, a possible candidate is the low frequency band of kinetic GAM, with the theoretical frequency of $f^{theo}_{GAM}=0.2\sqrt{1-1.4/q^2}Cs/(2\pi qR_0)$ \cite{Gao2006POP,Gao2008POP}. However, the low frequency GAM is only the theoretical solution considering the finite-gyroradius effect and has never been reported experimentally. For the I-mode edge plasmas with $q=4-6$ at the ETRO location, $f^{theo}_{GAM}$ is more than an order of magnitude smaller than the standard GAM, also inconsistent with the experimental frequency.


\begin{figure}[htbp]
\centering
	\includegraphics[width=0.8\textwidth,clip]{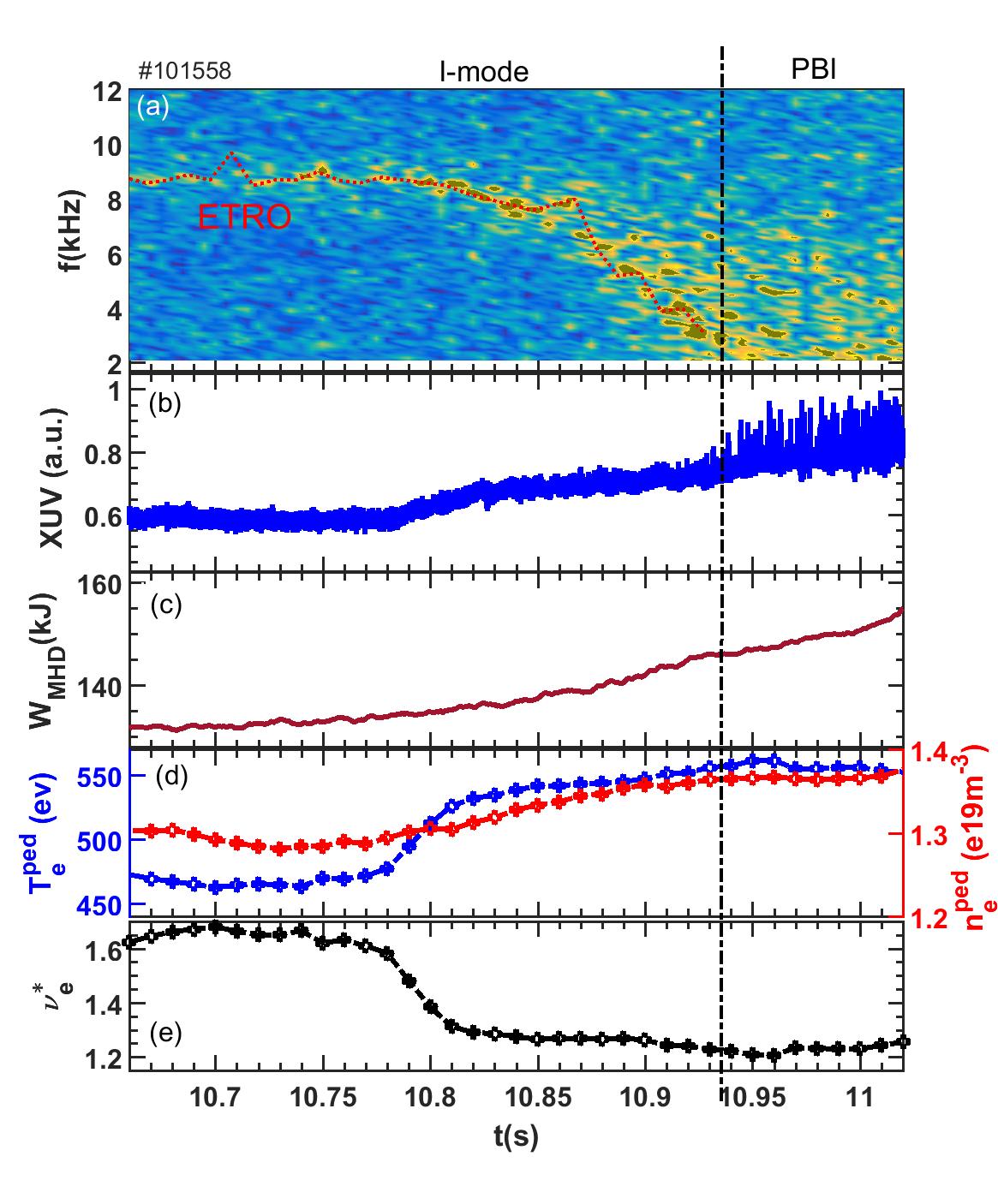}\\
	\caption{\label{f14} Temporal evolutions of (a) ETRO spectrum, (b) XUV signal, (c) $W_{MHD}$, (d) $T_e$ and $n_e$, (e) $\nu*_{e}$ at the pedestal top during I-mode to H-mode transition in shot $101558$.}
\end{figure}

To further reveal the nature of ETRO, the temporal evolution of ETRO during I-mode to H-mode transition is shown in figure \ref{f14}, as well as the radiation signal, electron density, electron temperature, and the normalized electron collisionality  $\nu*_{e}$, which is given by $\nu*_{e}=6.921\times10^{-18}q_{95}Rn_eZ_{eff}\ln\Lambda/(T_e^2\epsilon^{3/2})$, with major radius $R$, effective ion charge $Z_{eff}$, Coulomb logarithm $\ln\Lambda$ and inverse aspect ratio $\epsilon$ \cite{Sauter1999POP}. Here PBI means pedestal burst instability, which is the transition stage between stationary I-mode and H-mode and has been investigated in detail previously \cite{Zhong2022NF}. It could be seen that as close to the PBI stage, ETRO frequency decreased and reached about $2kHz$ finally. It should be noted that pedestal electron temperature $T^{ped}_e$ is certainly increased, suggesting that ETRO would not be the category of ion acoustic model. The consistency between decreased $\nu*_{e}$ and ETRO frequency suggested that ETRO is probably the stationary zonal flow with finite frequency, especially considering that it ended up with frequency close to LCO \cite{XuPRL2011}. However, it should be noted that after $t=10.85s$ $\nu*_{e}$ is nearly constant while ETRO frequency decreased more quickly, suggesting that other important factors have not been identified. The issue needs further investigation in the future.

\section{SMBI fueling effect on ETRO}
\begin{figure}[htbp]
\centering
	\includegraphics[width=0.8\textwidth,clip]{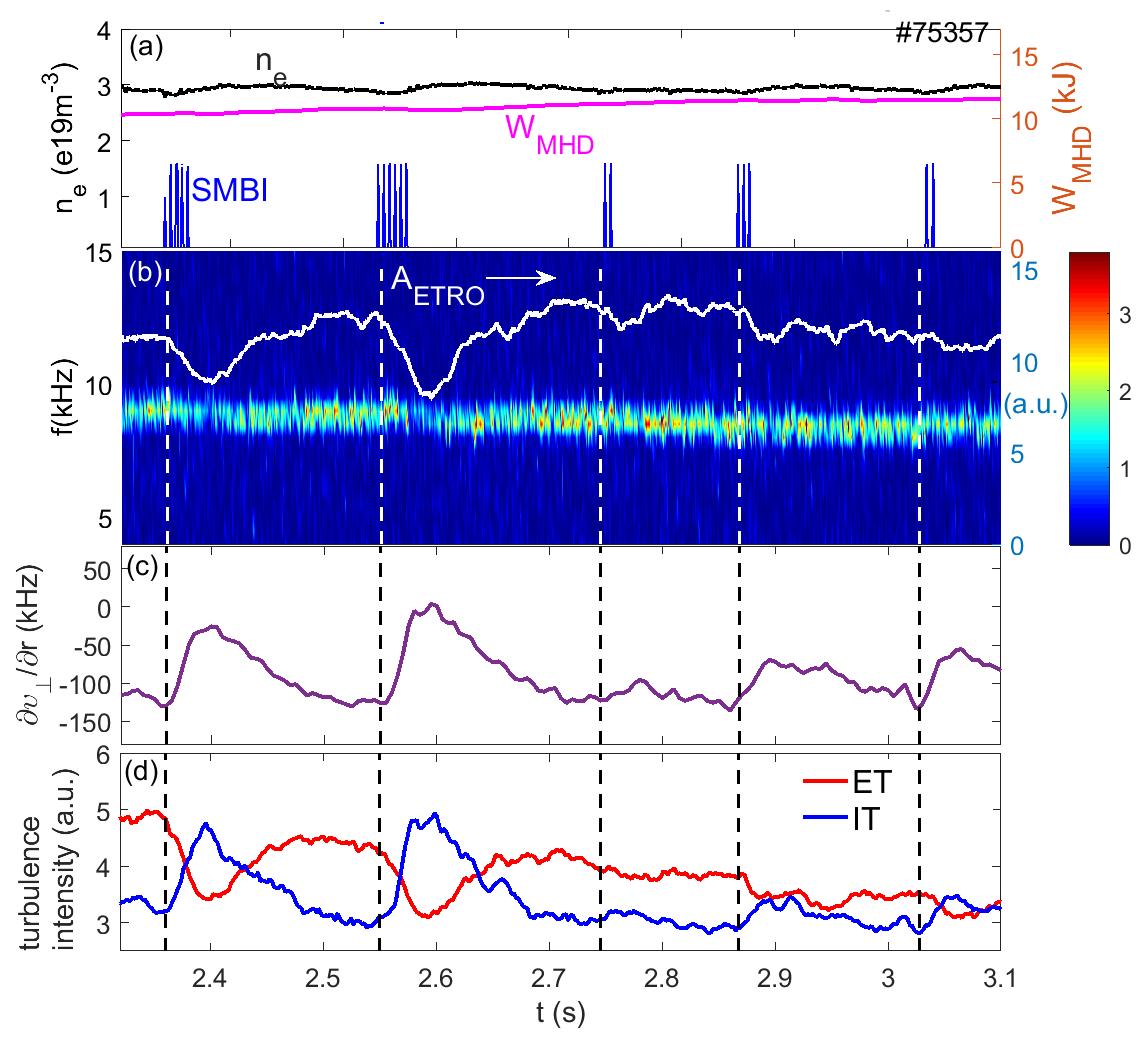}\\
	\caption{\label{f15} Temporal evolutions of (a) ETRO spectrum and intensity, (b) chord-averaged density, SMBI and $W_{MHD}$ signals, (c) rotation velocity shear and (d) ET/IT turbulence intensity in I-mode shot $75357$.}
\end{figure}

Supersonic molecular beam injection (SMBI) \cite{Yao2007NF} is used for fueling and density feedback control in EAST. It was found that modest SMBI could affect ETRO intensity without interrupting the I-mode confinement. Figure \ref{f15} displayed the evolutions of ETRO and IT/ET intensities, as well as the corresponding rotation velocity shear $\partial\upsilon_{\perp}/\partial r$ during stationary I-mode in shot $75357$. Here SMBI was used to maintain the chord-averaged density at the set value, and the number of SMBI pulses is different, ranging from $2$ to $6$ with each pulse interval of $2ms$. From fig.\ref{f15}(b) it could be seen that ETRO intensity would decrease obviously when SMBI starts, while the more the pluse number, the more ETRO intensity decreases. Fig.\ref{f15}(c) displayed the evolution of $\upsilon_{\perp}$ shear from $\upsilon_{\perp}$ radial profile. It should be noted that the shear is estimated from the inside of $\upsilon_{\perp}$ well, which is where ETRO appears \cite{Liu_2020,Zhong2022NF}. The velocity shear  $\partial\upsilon_{\perp}/\partial r$ is significantly damped by SMBI injection, which could be attributed to the neoclassical poloidal flow damping due to the edge plasma temperature decreasing, consistent with the results on other Tokamak devices \cite{Xiao2012NF,Zhong2020NF}. 

The most interesting result is the opposite evolutions of IT and ET as SMBI injection, as shown in fig.\ref{f15}(d). Here it is comprehensible that IT enhancement is due to the weakened flow shear.
While ET decrease is probably due to the decrease of edge plasma temperature. As illustrated by the simulation in section $4$, ET is most likely driven by $\nabla T_e$ in pedestal region, and the injecting neutral particles from SMBI would directly mitigate the driving source of ET. The decrease of ETRO intensity is the combination of ET decrease and IT enhancement. It should be noted that at $t=2.74s$, even two pulses could slightly influence ETRO and IT/ET intensities, further suggesting that SMBI could be an effective tool to control ETRO and turbulence in the I-mode edge plasma region.

\section{Conclusion and Discussion}
Considering that ETRO may play the key role for sustaining the stationary I-mode in EAST and such low frequency oscillation with azimuthally symmetric structure was also reported during I-mode in other devices, the main characteristics of ETRO were investigated and compared with that of GAM:

1) The toroidal symmetry of ETRO could be confirmed from the 16-group toroidal magnetic probes (fig.\ref{f4}), while the poloidal symmetry could be confirmed from the coherent analyses between XUV array and DR signal (fig.\ref{f5}), or through the radiation reconstruction. Direct evidences for the symmetry of potential fluctuations of ETRO is still not obtained. Considering that the pressure sideband of GAM should have an $m/n=1/0$ up-down asymmetry \cite{Diamond2005PPCF,Conway2021NF}, ETRO is unlikely to be GAM, but more like the symmetry modulation of pressure profile during the I-phase. 

2) During L-mode to I-mode transition with edge $T_e$ and $T_i$ both increasing, ETRO would appear with a smaller frequency than GAM in L-mode (fig.\ref{f7}), suggesting that ETRO is at least not a successor to the L-mode GAM;

3) ETRO has distinct harmonics in various diagnostics (fig.\ref{f8}) while harmonics of the $\upsilon_{\perp}$ spectrum is caused by amplitude asymmetry due to different phase velocities of IT-ET turbulence transition. On the other hand, the harmonics of GAM were only reported in rare cases \cite{Nagashima2007PPCF,Horvath2016NF,Qiu2017POP}.

4) The poloidal magnetic component of ETRO is dominated by $m=1$ structure (fig.\ref{f9}), similar to the M-mode in JET and I-phase in AUG \cite{Refy2020}, but different with the typical $m=2$ structure of GAM magnetic component. 

5) The radial propagation is also significantly different between ETRO and GAM. ETRO decayed radially outward much more quickly than inward (fig.\ref{f10}), while mostly GAM would prefer to propagate outward.

6) The most important property of ETRO is the accompanying turbulence transition between IT and ET (fig.\ref{f11}), and preliminary simulation suggested that ET is probably TEM and IT is probably ITG. This is much different from the modulation coupling between GAM and turbulence. 

7) Although statistics of the discharges with GAM and ETRO coexistence show that ETRO frequency indeed has a weakly positive correlation with $T^{ped}_e$, ETRO frequency is usually $2-3$ times smaller than that of GAM (fig.\ref{f13}), which could not be explained by current theoretical dispersion relation of kinetic GAM.

8) ETRO frequency would decrease rapidly with increasing $T^{ped}_e$ and decreasing $\nu*_{e}$ as I-mode approaching to H-mode (fig.\ref{f14}), implying that collisional damping is an important ingredient during the I-H transition.

In summary, most features of ETRO do not agree with GAM. It is probably a novel kind of finite frequency zonal flows, and unique to I-mode. The ETRO frequency is always several times larger than LCO during I-phase, implying that besides collisional damping the turbulence transition process may induce other nonlinear effects, which would be further investigated in the near future. 


\section{Acknowledgments}
 The authors thanks the fruitful discussions with Prof. Guo Zhibin. This work was supported in part by Natural Science Foundation of China under Grant Nos.U1967206, 11975231 and 11922513, National MCF Energy R$\&$D Program under Grant Nos. 2022YFE03010003 and 2018YFE0311200, the Collaborative Innovation Program of Hefei Science Center CAS No.2022HSC-CIP023, and the Fundamental Research Funds for the Central Universities WK3420000018.
\section*{Appendix}
EAST I-mode working group: 
\\
Y.T. Song$^1$, X.L. Zou$^2$, E.Z. Li$^1$, B. Zhang$^1$, Y. Chao$^1$, F. Chen$^1$, X.Y. Chen$^3$,  Y.X. Cheng$^1$, Y.Q. Chu$^1$, B.J. Ding$^1$, R. Ding$^1$, Y.M. Duan$^1$, Y.F. Jin$^1$, S.Y. Lin$^1$, A.D. Liu$^3$, H.Q. Liu$^1$,   B. Lv$^1$, L.Y. Meng$^1$, C.M. Qin$^1$, H.L. Wang$^1$, L. Wang$^1$, P. Wang$^1$, S.X. Wang$^1$, X.J. Wang$^1$, Y.F. Wang$^1$, C.B. Wu$^1$, T.Y. Xia$^1$, L.Q. Xu$^1$, G.H. Yan$^1$, Y. Yang$^1$, L. Yu$^1$,  F.L. Zhang$^1$, L. Zhang$^1$, X.J. Zhang$^1$, X.M. Zhong$^3$, R.J. Zhu$^3$.
\\
{$^1$ Institute of Plasma Physics, Chinese Academy of Sciences, Anhui Hefei 230021, China}
\\
{$^2$ CEA, IRFM, F-13108 St Paul Les Durance, France}
\\
{$^3$ School of Nuclear Science and Technology, University of Science and Technology of China, Anhui Hefei 230026, China}
\section*{References}
\bibliographystyle{unsrt}
\bibliography{ETRO}

\end{document}